# Intrinsic current drive by electromagnetic electron drift wave turbulence in tokamak pedestal region

Wen He[1], Lu Wang*[1] and Ge Zhuang[2]

[1]International Joint Research Laboratory of Magnetic Confinement Fusion and Plasma Physics, State Key Laboratory of Advanced Electromagnetic Engineering and Technology, School of Electrical and Electronic Engineering, Huazhong University of Science and Technology, Wuhan, 430074, China

[2]Department of Engineering and Applied Physics School of Physical Sciences, University of Science and Technology of China, Hefei, 230026, People's Republic of China.

*Author to whom any correspondence should be addressed, E-mail: luwang@hust.edu.cn

**Abstract:** The local intrinsic parallel current density driven by electron drift wave (DW) turbulence including electromagnetic (EM) effects is analytically studied. The scalings of the ratios of intrinsic current density driven by residual turbulent flux and by turbulent source to the bootstrap current density with electron density and temperature are predicted to be $T_e^{3/4}T_i/n_e$ and $T_e T_i/n_e$, respectively. Based on the typical parameters in DIII-D pedestal region, the local intrinsic current density driven by both the residual turbulent flux and the turbulent source is negligible. However, despite the negligible turbulent source driven current, the residual turbulent flux driven local intrinsic current density by EM DW turbulence can reach about 66% of the bootstrap current density for ITER pedestal parameters due to much lower collisionality in ITER than in DIII-D. Moreover, the contributions from adiabatic ES parts, non-adiabatic ES parts and non-adiabatic EM parts of the plasma response to electromagnetic fluctuations are analyzed. It is found that there exists strong cancelation between non-adiabatic ES response and the non-adiabatic EM response for the ITER pedestal case, and thus the kinetic stress contributed by the adiabatic ES response of parallel electron pressure dominates the intrinsic current drive. This is different from the ES electron DW case. Therefore, the EM effects on turbulence driven intrinsic current density should be carefully considered in the future reactor with high ratio of electron pressure to the magnetic pressure and steep pressure profile.

# 1. Introduction

Current density profile is closely related to magnetohydrodynamic (MHD) instability in tokamaks, such as the neoclassical tearing mode (NTM) in the core region and edge localized mode (ELM) in the pedestal region. The neoclassical bootstrap (BS) current has attracted wide attention [1-4]. This is not only because it provides an economic way of driving the plasma current, but also it has strong impact on the MHD instability. It has been discovered that the ubiquitous turbulence in tokamaks can also affect the current density profile via hyper-resistivity and anomalous resistivity [5, 6]. Moreover, by making an analogy between the collisional scattering and the resonant electron scattering by drift wave (DW) turbulence, the turbulence-driven BS current has been proposed [7]. It has been found that the onset threshold of NTM can be significantly affected by the turbulence or turbulence driven current density [8, 9]. The interaction between micro-turbulence and tearing mode has been investigated via both self-consistent theoretical model [10, 11] and Landau-fluid simulations [12, 13]. These may be relevant to the effects of micro-turbulence on the onset and recovery of NTM observed on JT-60U and DIII-D [14, 15]. Therefore, the study of intrinsic current density driven by micro-turbulence is helpful for comprehensive understanding of the multi-scale interaction between micro-turbulence and MHD instability.

The idea of intrinsic current driven by electrostatic (ES) electron DW in sheared magnetic field in the presence of radial variation of fluctuation level was proposed in [16]. The intrinsic current density induced by finite value of averaged $k_\parallel$ (parallel wave number of the DW) is very similar to the intrinsic rotation driven by residual stress due to $k_\parallel$ symmetry breaking. Later, intrinsic current drive by various types of micro-turbulence in tokamak, e.g., electron temperature gradient (ETG) [17, 18], ion temperature gradient (ITG) [19] and trapped electron mode (TEM) [20] has been investigated. In these works, ES turbulence results in the change of local current density profile at least 10% as compared to the BS current density profile without considering electromagnetic (EM) effects. However, in pedestal region where the pressure profile is very steep so that $\hat{\beta} = \beta(qR/L_\perp)^2$ can be above unity with $\beta$ being the ratio of thermal pressure to magnetic pressure, $q$ being the safety factor, $R$ being the major radius and $L_\perp$ being the mean pressure profile scale length across the magnetic flux surfaces, the EM effects on micro-turbulence is of great importance [21]. Hence, the EM effects on turbulence driven current should be carefully considered in pedestal region. Although there exist some works on EM turbulence driven current density [22-24], a general model with explicit contributions from ES effects and EM effects using the relationship between ES potential and magnetic vector potential fluctuations is missing. The intrinsic current driven by ETG turbulence including explicit EM effects has been studied recently [25], where the estimation is based on the parameters of tokamak core plasmas. Therefore, the goal of this work is to study the intrinsic current driven by EM electron DW turbulence in pedestal region.

In this paper, the intrinsic current density driven by EM electron DW turbulence is estimated. The ratios of turbulence driven intrinsic current density from quasi-linear estimation to the BS current density based on the DIII-D and International Thermonuclear Experimental Reactor (ITER) pedestal parameters are presented. The intrinsic current density driven by EM DW turbulence may be too small to affect the current density profile in DIII-D pedestal region. However, the local intrinsic current density driven by residual turbulent flux can reach about 66% of the BS current density in ITER pedestal due to much lower collisionality. Moreover, an important founding is that the non-adiabatic EM effects strongly cancels the non-adiabatic ES effects, thus the dominate contribution to intrinsic current drive comes from the adiabatic ES response of parallel electron pressure induced kinetic stress. This indicates the necessity of considering the EM effects on intrinsic current driven by turbulence in future reactor with relatively high $\beta$ and steep pressure profile region.

The remainder of this paper is organized as follows. In section 2, a detailed quasi-linear derivation of the residual turbulent flux and the turbulent source driven by EM electron DW turbulence is presented. Besides, based on DIII-D and ITER pedestal parameters, the ratio of intrinsic current density to the BS current density is estimated. Finally, a summary and some discussions are given in section 3.

## 2. Quasi-linear estimate for intrinsic parallel current density

We will use the mean parallel current density evolution equation derived from the EM gyro-kinetic equation [25],

$$\frac{\partial \langle J_\parallel \rangle}{\partial t} + \nabla \cdot \langle \delta \boldsymbol{v}_{E \times B, r} \delta J_\parallel \rangle - \nabla \cdot \langle \frac{e}{m_e} \delta \boldsymbol{b}_r \delta P_\parallel \rangle$$
$$= \langle -\frac{e^2}{m_e} \hat{\boldsymbol{b}} \cdot \nabla \delta \phi \delta n_e \rangle + \langle -\frac{e^2}{c m_e} \frac{\partial \delta A_\parallel}{\partial t} \delta n_e \rangle. \tag{1}$$

Here, $J_\parallel$ is the parallel current density, $\delta \boldsymbol{v}_{E \times B} = \frac{c \hat{\boldsymbol{b}} \times \nabla \delta \phi}{B}$ is the fluctuating $\boldsymbol{E} \times \boldsymbol{B}$ drift velocity, $\hat{\boldsymbol{b}} = \boldsymbol{B}/B$ is the unit vector of equilibrium magnetic line, $\delta n_e = \int \delta f d^3 v$ is the perturbed electron density with $\delta f$ being the perturbed electron distribution function, $\delta J_\parallel = -e \int \delta f v_\parallel d^3 v$ is the perturbation of parallel current density, $\delta P_\parallel = m_e \int \delta f v_\parallel^2 d^3 v$ is the perturbation of the parallel electron pressure, $\delta \phi$ is the ES potential fluctuation, $c$ is the speed of light, $m_e$ is the electron mass, $e$ is the elementary charge, $\delta \boldsymbol{b}_\perp = \frac{\delta \boldsymbol{B}_\perp}{B}$ is the normalized fluctuating perpendicular magnetic field where $\delta \boldsymbol{B}_\perp \approx -\hat{\boldsymbol{b}} \times \nabla \delta A_\parallel$ is the perturbed perpendicular magnetic field. We only consider the shear component of magnetic perturbation, i.e., $\delta A_\parallel$, and the compressional component $\delta B_\parallel$ is not included. The $\langle \cdots \rangle$ in this paper represents the flux average. On the left hand side (LHS), the terms under the divergence are turbulent flux $\Gamma_r$ with $\langle \delta v_{E \times B, r} \delta J_\parallel \rangle$ being the Reynolds stress-like term and $-\langle \frac{e}{m_e} \delta b_r \delta P_\parallel \rangle$ being the kinetic stress-like term. On the right hand side (RHS), the turbulent source are driven by the correlation between density and parallel electric field fluctuations including both the electrostatic field $S_1 = \langle -\frac{e^2}{c m_e} \hat{\boldsymbol{b}} \cdot \nabla \delta \phi \delta n_e \rangle$ and the inductive electric field $S_2 = \langle -\frac{e^2}{c m_e} \frac{\partial \delta A_\parallel}{\partial t} \delta n_e \rangle$. In this work, we focus on the intrinsic current drive which is independent of the mean parallel current density or its gradient. This is analogous to the intrinsic rotation drive by the residual stress, kinetic stress, and turbulent source [26-28].

## A. Residual turbulent flux and turbulent source

The quasi-linear estimation for the residual turbulent flux and turbulent source requires the calculation of linear response of $\delta n_e$, $\delta J_\parallel$ and $\delta P_\parallel$ to the electromagnetic fluctuations. By linearizing the EM drift equation of electrons, the linearized perturbed electron distribution function in Fourier space decomposed into adiabatic and non-adiabatic parts can be written as,

$$\delta f_k = \frac{e\delta\phi_k}{T_e} F_M + \delta g_k$$

$$= \frac{e\delta\phi_k}{T_e} F_M - n_e \left(\frac{m_e}{2\pi T_e}\right)^{3/2} \exp(-\hat{v}_\parallel^2 - \hat{v}_\perp^2) \left\{\omega_k - \omega_{*e}\left[1 + \eta_e\left(\hat{v}_\parallel^2 + \hat{v}_\perp^2 - \frac{3}{2}\right)\right]\right\} \frac{e}{T_e}\left(\delta\phi_k - \frac{v_\parallel \delta A_{\parallel k}}{c}\right)\frac{1}{\omega_k - \omega_t}. \quad (2)$$

Here, $F_M$ is the Maxwellian distribution function of electrons, $\delta g_k$ represents the non-adiabatic part of the perturbed electron distribution function, $\omega_{*e} = \frac{cT_e}{-eB}\hat{\boldsymbol{b}} \times \nabla\ln n_e \cdot \boldsymbol{k}$ is the diamagnetic drift frequency of electron with $\boldsymbol{k}$ being the wave vector, $\omega_t = k_\parallel \hat{v}_\parallel v_{the}$ is the electron transit frequency with $k_\parallel$ being the parallel wave number, $v_{the} = \sqrt{\frac{2T_e}{m_e}}$ is the electron thermal velocity with $T_e$ being the electron temperature, $\hat{v}_\parallel = \frac{v_\parallel}{v_{the}}$ is the normalized electron parallel velocity, $\hat{v}_\perp = \frac{v_\perp}{v_{the}}$ is the normalized electron perpendicular velocity, $\eta_e = L_n/L_{Te}$ with $L_{Te} = -(\nabla\ln T_e)^{-1}$ and $L_n = -(\nabla\ln n_e)^{-1}$ being the electron temperature gradient length scale and electron density gradient length scale, respectively. It should be pointed out that only the transit resonance will be considered in this work.

Then, taking the moments of the linearized perturbed electron distribution function, the linear response $\delta n_e$, $\delta J_\parallel$ and $\delta P_\parallel$ can be obtained. The perturbed electron density is

$$\delta n_{ek} = \frac{e\delta\phi_k}{T_e} n_e + \int \delta g_k d^3v$$

$$= n_e \delta\hat{\phi}_k - \sqrt{\pi} n_e \exp(-\xi_e^2) \frac{\gamma_k}{|k_\parallel| v_{the}} \left(\delta\hat{\phi}_k - 2\xi_e \delta\hat{A}_{\parallel k}\right)$$

$$+i\sqrt{\pi}n_e\exp(-\xi_e^2)\frac{1}{|k_\parallel|v_{\text{the}}}\left\{\omega_{kr}-\omega_{*e}\left[1+\eta_e\left(\xi_e^2-\frac{1}{2}\right)\right]\right\}(\delta\hat{\phi}_k$$

$$-2\xi_e\delta\hat{A}_{\parallel k}). \tag{3}$$

Here, $\delta\hat{\phi}_k=\frac{e\delta\phi_k}{T_e}$ and $\delta\hat{A}_{\parallel k}=\frac{\delta A_{\parallel k}}{\rho_e B}$ are normalized fluctuating potentials with $\rho_e=\frac{v_{\text{the}}}{\Omega_e}$ being the gyro-radius of electron, $\Omega_e=\frac{eB}{cm_e}$ being the gyro-frequency of electron. $\omega_k=\omega_{kr}+i\gamma_k$ is taken with $\omega_{kr}$ being the real frequency and $\gamma_k$ being the linear growth rate, respectively. Especially, $\gamma_k^2\ll\omega_{kr}^2$ is assumed and $\xi_e=\frac{\omega_{kr}}{k_\parallel v_{\text{the}}}$ is defined. The first term on the RHS of Eq. (3) comes from the adiabatic response ($\delta n_{ekES}^A$) to ES potential and the other terms come from the non-adiabatic response. Besides, the non-adiabatic response includes both the ES and EM parts ($\delta n_{ekES}^{NA}$ and $\delta n_{ekEM}^{NA}$) corresponding to the terms proportional to $\delta\hat{\phi}_k$ and $\delta\hat{A}_{\parallel k}$, respectively. Then, Eq. (3) can be rewritten as

$$\delta n_{ek}=\delta n_{ekES}^A+\delta n_{ekES}^{NA}+\delta n_{ekEM}^{NA}. \tag{4}$$

The perturbed parallel current is

$$\delta J_{\parallel k}=\sqrt{\pi}en_e v_{\text{the}}\xi_e\exp(-\xi_e^2)\frac{\gamma_k}{|k_\parallel|v_{\text{the}}}(\delta\hat{\phi}_k-2\xi_e\delta\hat{A}_{\parallel k})$$

$$-i\sqrt{\pi}en_e v_{\text{the}}\xi_e\exp(-\xi_e^2)\frac{1}{|k_\parallel|v_{\text{the}}}\{\omega_{kr}$$

$$-\omega_{*e}\left[1+\eta_e\left(\xi_e^2-\frac{1}{2}\right)\right]\}(\delta\hat{\phi}_k-2\xi_e\delta\hat{A}_{\parallel k}). \tag{5}$$

It is noted that the perturbed parallel current is produced by the non-adiabatic response, since the adiabatic ES response does not have contribution. Similarly, on the RHS of Eq. (5), the terms relevant to $\delta\hat{\phi}_k$ and $\delta\hat{A}_{\parallel k}$ are resulted from the non-adiabatic ES and EM response, respectively. Then, Eq. (5) can be rewritten as

$$\delta J_{\parallel k}=\delta J_{\parallel kES}^{NA}+\delta J_{\parallel kEM}^{NA}. \tag{6}$$

The perturbed parallel electron pressure is

$$\delta P_{\parallel k} = P_{\parallel}\delta\hat{\phi}_k - 2\sqrt{\pi}P_{\parallel}\xi_e^2\exp(-\xi_e^2)\frac{\gamma_k}{|k_{\parallel}|v_{\text{the}}}\left(\delta\hat{\phi}_k - 2\xi_e\delta\hat{A}_{\parallel k}\right)$$

$$+ i2\sqrt{\pi}P_{\parallel}\xi_e^2\exp(-\xi_e^2)\frac{1}{|k_{\parallel}|v_{\text{the}}}\left\{\omega_{kr} - \omega_{*e}\left[1 + \eta_e\left(\xi_e^2 - \frac{1}{2}\right)\right]\right\}(\delta\hat{\phi}_k$$

$$- 2\xi_e\delta\hat{A}_{\parallel k}). \tag{7}$$

Here, the first term on the RHS comes from the adiabatic ES response. The other terms come from the non-adiabatic ES and EM responses, again. Then, Eq. (7) can be rewritten as

$$\delta P_{\parallel k} = \delta P_{\parallel k\text{ES}}^{\text{A}} + \delta P_{\parallel k\text{ES}}^{\text{NA}} + \delta P_{\parallel k\text{EM}}^{\text{NA}}. \tag{8}$$

Based on Eqs. (3)-(8), the calculations of residual turbulent flux and turbulent source are straightforward.

The detailed calculations of $\langle\delta v_{E\times B,r}\delta J_{\parallel}\rangle$ and $\langle-\frac{e}{m_e}\delta b_r\delta P_{\parallel}\rangle$ can be found in Appendix. The expression of residual turbulent flux can be written as

$$\Gamma_r = \sum_k \frac{1}{2}\sqrt{\pi}k_\theta\rho_e en_e v_{\text{the}}^2\xi_e\exp(-\xi_e^2)\left\{\frac{\omega_{kr}}{|k_{\parallel}|v_{\text{the}}}\right.$$

$$\left. - \frac{\omega_{*e}}{|k_{\parallel}|v_{\text{the}}}\left[1 + \eta_e\left(\xi_e^2 - \frac{1}{2}\right)\right]\right\}\left(|\delta\hat{\phi}_k|^2 - 4\xi_e\text{Re}\langle\delta\hat{A}_{\parallel k}\delta\hat{\phi}_{-k}\rangle\right.$$

$$\left. + 4\xi_e^2|\delta\hat{A}_{\parallel k}|^2\right) + \sum_k \frac{1}{2}k_\theta\rho_e en_e v_{\text{the}}^2\text{Im}\langle\delta\hat{A}_{\parallel k}\delta\hat{\phi}_{-k}\rangle. \tag{9}$$

Here, the first term on the RHS comes from pure ES contribution from $\delta J_{\parallel k\text{ES}}^{\text{NA}}$ to Reynold stress. The other terms come from EM effects including both the contribution from non-adiabatic EM response of current $\delta J_{\parallel k\text{EM}}^{\text{NA}}$ to the Reynold stress and the contribution from non-adiabatic ES response of electron pressure $\delta P_{\parallel k\text{ES}}^{\text{NA}}$ to the kinetic stress. Especially, the last term comes from the contribution from adiabatic response of parallel electron pressure $\delta P_{\parallel k\text{ES}}^{\text{A}}$ to the kinetic stress. If the ES limit is taken, Eq. (9) will be reduced to the same form as Eq. (6) in [20]. In the same way, the expression of turbulent source is

$$S = -\sum_k \frac{\sqrt{\pi}}{2} e n_e v_{\text{the}}^2 \exp(-\xi_e^2) \left\{ \frac{\omega_{kr}}{|k_\parallel| v_{\text{the}}} \right.$$

$$\left. - \frac{\omega_{*e}}{|k_\parallel| v_{\text{the}}} \left[1 + \eta_e \left(\xi_e^2 - \frac{1}{2}\right)\right] \right\} \left( k_\parallel |\delta\hat{\phi}_k|^2 - \frac{4\omega_{kr}}{v_{\text{the}}} \text{Re}\langle\delta\hat{A}_{\parallel k} \delta\hat{\phi}_{-k}\rangle \right.$$

$$\left. + \frac{2\gamma_k}{v_{\text{the}}} \text{Im}\langle\delta\hat{A}_{\parallel k} \delta\hat{\phi}_{-k}\rangle + \frac{4\xi_e \omega_{kr}}{v_{\text{the}}} |\delta\hat{A}_{\parallel k}|^2 \right)$$

$$- \sum_k e n_e v_{\text{the}} \left[ \omega_{kr} \text{Im}\langle\delta\hat{A}_{\parallel k} \delta\hat{\phi}_{-k}\rangle + \gamma_k \text{Re}\langle\delta\hat{A}_{\parallel k} \delta\hat{\phi}_{-k}\rangle \right]. \quad (10)$$

The detailed calculations of $S$ can be also found in Appendix. The first term on the RHS of Eq. (10) comes from the pure ES contribution, i.e., correlation between the non-adiabatic ES electron density response $\delta n_{ekES}^{NA}$ and the ES electric field fluctuation, which is consistent with the result in [22]. Here, it is noted that the adiabatic ES density response $\delta n_{ekES}^{A}$ does not contribute to the pure ES turbulent source. The other terms are due to EM effects including the correlation between the non-adiabatic EM density response $\delta n_{ekEM}^{NA}$ and both the ES and inductive electric field fluctuations and the correlation between the adiabatic ES density response $\delta n_{ekES}^{A}$ and the inductive electric field fluctuation.

## B. Intrinsic current density and comparison with BS current density

Up to now, the quasilinear expression for the residual turbulent flux and the turbulent source have been written in terms of the EM fluctuations. Then, the explicit estimation of EM effects on intrinsic current drive requires the relation between $\delta\hat{A}_{\parallel k}$ and $\delta\hat{\phi}_k$. Combing Eq. (5) with the Ampere's law, $-\nabla^2 \delta A_\parallel = \frac{4\pi}{c} \delta J_\parallel$, the relation can be written as

$$\delta\hat{A}_{\parallel k} = \frac{C_1 - iC_2}{D} \delta\hat{\phi}_k. \quad (11)$$

Here, $C_1 = F_1 \frac{k_\perp^2 \rho_e^2}{\beta_e} + 2\xi_e F_1^2 + 2\xi_e F_2^2$, $C_2 = F_2 \frac{k_\perp^2 \rho_e^2}{\beta_e}$, $D = \left(\frac{k_\perp^2 \rho_e^2}{\beta_e} + 2\xi_e F_1\right)^2 + 4\xi_e^2 F_2^2$ with $F_1 = \sqrt{\pi} \xi_e \exp(-\xi_e^2) \frac{\gamma_k}{|k_\parallel| v_{\text{the}}}$ and $F_2 = \sqrt{\pi} \xi_e \exp(-\xi_e^2) \left\{ \frac{\omega_{kr}}{|k_\parallel| v_{\text{the}}} - \frac{\omega_{*e}}{|k_\parallel| v_{\text{the}}} \left[1 + \eta_e \left(\xi_e^2 - \frac{1}{2}\right)\right] \right\}$, are all dimensionless. $k_\perp$ is perpendicular wave number, and $\beta_e = \frac{n_e T_e}{8\pi B^2}$ is the ratio between electron pressure and the magnetic pressure. When the finite

Larmor radius effects are neglected, Eq. (11) yields to $\delta\hat{A}_{\|k} = \frac{1}{2}\frac{k_\|v_{the}}{\omega}\delta\hat{\phi}_k$. Note that $\delta\hat{A}_{\|k} = \frac{\delta A_{\|k}}{\rho_e B}$ and $\delta\hat{\phi}_k = \frac{e\delta\phi_k}{T_e}$ are normalized, so the ideal Alfven wave limit $\delta A_{\|k} = \frac{k_\|c}{\omega}\delta\phi_k$, i.e., $\delta E_\| = 0$ can be reproduced. In this limit, the turbulent source which is proportional to $\delta E_\|$ vanishes. The non-adiabatic responses in Eqs. (3), (5) and (7) are all proportion to $(\delta\hat{\phi}_k - 2\xi_e\delta\hat{A}_{\|k})$. From Eq. (11), we can obtain

$$\delta\hat{\phi}_k - 2\xi_e\delta\hat{A}_{\|k} = \frac{\frac{k_\perp^4\rho_e^4}{\beta_e^2} + 2\xi_eF_1\frac{k_\perp^2\rho_e^2}{\beta_e} + i2\xi_eF_2\frac{k_\perp^2\rho_e^2}{\beta_e}}{\left(\frac{k_\perp^2\rho_e^2}{\beta_e} + 2\xi_eF_1\right)^2 + 4\xi_e^2F_2^2}\delta\hat{\phi}_k. \tag{12}$$

Substituting Eq. (12) into the residual turbulent flux and turbulent source, it is found that $k_\|$ symmetry breaking is required for non-zero intrinsic current drive. This is very similar to the intrinsic rotation drive. Various symmetry breaking mechanisms have been proposed, such as $\boldsymbol{E}\times\boldsymbol{B}$ shear [29, 30], intensity gradient [31], and so on [32, 33]. Symmetry breaking induced by $\boldsymbol{E}\times\boldsymbol{B}$ shear will be taken in this work. The average of parallel wave number induced by the $\boldsymbol{E}\times\boldsymbol{B}$ shear for electron DW turbulence is $\bar{k}_\| = \sum_{\vec{k}}k_\||\delta\hat{\phi}_k|^2/\sum_{\vec{k}}|\delta\hat{\phi}_k|^2 \simeq \frac{q}{2\hat{s}}k_\theta\rho_i\frac{\rho_i}{RL_p}$ [22] with $L_p$ being the length scale of pressure gradient. Considering $\omega_{*e} = \frac{k_\theta\rho_e v_{the}}{2L_n}$ and $I_k = |\delta\hat{\phi}_k|^2$, then substituting Eq. (12) to Eq. (9), the residual turbulent flux can be written as

$$\Gamma_r = \frac{1}{2}en_ev_{the}^2\left[\left(1 - 4\xi_e\frac{C_1}{D}\right.\right.$$
$$+ 4\xi_e^2\frac{C_1^2 + C_2^2}{D^2}\bigg)H\xi_e\frac{\omega_{*e}}{k_\|v_{the}}\frac{\omega_{kr}}{|k_\||v_{the}}L_n\frac{q}{\hat{s}}k_\theta\rho_i\frac{\rho_i}{RL_p}\sum_k I_k$$
$$\left.- \frac{C_2}{D}\frac{\omega_{*e}}{k_\|v_{the}}L_n\frac{q}{\hat{s}}k_\theta\rho_i\frac{\rho_i}{RL_p}\sum_k I_k\right], \tag{13}$$

where, $H = \sqrt{\pi}\exp(-\xi_e^2)\left\{1 - \frac{\omega_{*e}}{\omega_{rk}}\left[1 + \eta_e\left(\xi_e^2 - \frac{1}{2}\right)\right]\right\}$.

The turbulent source can be written as

$$S = -\frac{en_e v_{\text{the}}^2}{2R}\left[\left(\frac{1}{2} - 2\frac{C_1}{D}\xi_e - \frac{C_2}{D}\frac{\gamma_k}{k_\parallel v_{\text{the}}} + 2\xi_e^2\frac{C_1^2 + C_2^2}{D^2}\right)H\frac{\omega_{kr}}{|k_\parallel|v_{\text{the}}}\frac{q}{\hat{s}}k_\theta\rho_i\frac{\rho_i}{L_p}\sum_k I_k\right.$$

$$\left. - \left(\frac{C_2}{D}\xi_e - \frac{C_1}{D}\frac{\gamma_k}{k_\parallel v_{\text{the}}}\right)\frac{q}{\hat{s}}k_\theta\rho_i\frac{\rho_i}{L_p}\sum_k I_k\right]. \tag{14}$$

It should be mentioned that although both the turbulent flux and source seem to diverge as approaching to the rational surface, i.e., $|k_\parallel| \to 0$ ($\xi_e^2 \to \infty$), the factor of exponential convergence $\exp(-\xi_e^2)$ will regularize the magnitude of the turbulent flux and source. This has been discussed in Ref. 20. The transit resonance condition in cylindrical geometry, $\omega_t = \omega_{kr} \approx \omega_{*e}$ implies $\Delta x \approx \frac{L_s}{L_n}\sqrt{\frac{m_e}{m_i}}\rho_s$, where $L_s = qR/\hat{s}$ is the magnetic shear length with $\hat{s}$ being the magnetic shear. An interesting point is that the width of electron Landau layer, $\Delta x$ could be several ion gyroradii for normal magnetic shear due to very steep density gradient in pedestal region. This will result in the radial variation of the turbulent flux and source around the rational surface is slower than that in flat density gradient region [20]. In recent gyrokinetic simulation of CTEM turbulence driven current [34], the scale length of the corrugated current profile around the rational surface is about 5-10 gyroradii where the density profile is relatively steep and the magnetic shear is weak. This is consistent with our theory.

Then, the negative divergence of residual turbulent flux can provide a turbulent force for driving intrinsic current density. We take the length scale of variation of the residual turbulent flux as mesoscale, i.e., $\sqrt{\rho_s L_n}$, which is larger than the ion gyroradius but smaller than the density gradient scale length. Then, the residual turbulent flux driven force is $\mp\frac{\Gamma_r}{\sqrt{\rho_s L_n}}$. The sign of $\mp$ corresponds to positive (negative) gradient of turbulent flux. By balancing the turbulent flux driven force with the collisional friction force $-\nu_{ei}J_{\text{turb}}$, we can obtain the intrinsic current density driven by the residual turbulent flux

$$J_{\text{turb}}^\Gamma = \mp\frac{\Gamma_r}{\nu_{ei}\sqrt{\rho_s L_n}}. \tag{15}$$

Similarly, balancing the turbulent source $S$ with the collisional friction force produces the intrinsic current density driven by turbulent source $S$

$$J_{\text{turb}}^S = \frac{S}{\nu_{\text{ei}}}. \tag{16}$$

Now, we compare the intrinsic current density driven by EM electron DW turbulence with the BS current density. The BS current density can be estimated as following [22],

$$J_{\text{BS}} \simeq 5\sqrt{\frac{1}{\varepsilon}} \frac{cqn_e T_e}{BL_P}, \tag{17}$$

where $\varepsilon$ is the reverse aspect ratio. Subsequently, the ratios of the intrinsic current density driven by residual turbulent flux and turbulent source to the BS current density can be written as

$$\frac{J_{\text{turb}}^\Gamma}{J_{\text{BS}}} = \mp \frac{\sqrt{\varepsilon} v_{\text{the}}}{5\nu_{\text{ei}}\sqrt{\rho_s L_n}} \frac{L_n}{\hat{s}R} \frac{\rho_i}{\rho_e} \left[ \left(1 - 4\xi_e \frac{C_1}{D} \right. \right.$$
$$\left. + 4\xi_e^2 \frac{C_1^2 + C_2^2}{D^2} \right) H \xi_e \frac{\omega_{*e}}{k_\| v_{\text{the}}} \frac{\omega_{kr}}{|k_\|| v_{\text{the}}} k_\theta \rho_i \sum_k I_k$$
$$\left. - \frac{C_2}{D} \frac{\omega_{*e}}{k_\| v_{\text{the}}} k_\theta \rho_i \sum_k I_k \right], \tag{18}$$

and

$$\frac{J_{\text{turb}}^S}{J_{\text{BS}}} = -\frac{\sqrt{\varepsilon}}{5\nu_{\text{ei}}} \frac{v_{\text{the}}}{\hat{s}R} \frac{\rho_i}{\rho_e} \left[ \left(\frac{1}{2} - 2\frac{C_1}{D}\xi_e - \frac{C_2}{D}\frac{\gamma_k}{k_\| v_{\text{the}}} \right. \right.$$
$$\left. + 2\xi_e^2 \frac{C_1^2 + C_2^2}{D^2} \right) H \frac{\omega_{kr}}{|k_\|| v_{\text{the}}} k_\theta \rho_i \sum_k I_k$$
$$\left. - \left(\frac{C_2}{D}\xi_e - \frac{C_1}{D}\frac{\gamma_k}{k_\| v_{\text{the}}}\right) k_\theta \rho_i \sum_k I_k \right]. \tag{19}$$

We take typical pedestal parameters of DIII-D [35], $q = 3.6$, $R/L_{Te} = 144$, $R/L_n = 64$, $\hat{s} = 1$, $\varepsilon = 0.37$, $R = 1.77\text{m}$, $n_e = 2.48 \times 10^{19}/\text{m}^3$, $T_e = 197\text{eV}$, $T_i = 397\text{eV}$, $\nu_{\text{ei}} = 3.88 \times 10^5 \text{Hz}$, $\rho_e = 2.82 \times 10^{-5}\text{m}$ and $\beta_e = 0.07\%$. The typical EM electron DW turbulence scale in pedestal region is taken as $k_\theta \rho_s \simeq 0.28$ (with toroidal mode number being 30), $k_\perp^2 = 2k_\theta^2$, $\gamma_k/\omega_{*e} = 1/10$, $\sum_k I_k = 5 \times 10^{-4}$, $k_\|^2 = \frac{\hat{s}^2}{q^2 R^2}$ and $\omega_{kr} = \frac{\omega_{*e}}{2}$. We define $c_s = \sqrt{\frac{2T_e}{m_i}}$ and $\rho_s = \frac{c_s}{\Omega_i}$ where $\Omega_i = \frac{eB}{cm_i}$ is the Larmor frequency of ion. Then, the ratios can be obtained

$$\frac{J_{\text{turb}}^{\Gamma}}{J_{\text{BS}}} = \mp(0.58\% - 0.56\% + 2.32\% - 4.06\%), \tag{20}$$

and

$$\frac{J_{\text{turb}}^{S}}{J_{\text{BS}}} = -1.02\% + 9.98 \times 10^{-3} + 8.26 \times 10^{-4} - 4.00 \times 10^{-3}$$

$$+ (1.0\% - 0.24\%). \tag{21}$$

On the RHS of Eq. (20), the first term comes from the non-adiabatic ES contribution $\delta J_{\parallel k\text{ES}}^{\text{NA}}$ to $\langle \delta v_{E \times B,r} \delta J_{\parallel} \rangle$. While the second term includes the equivalent contributions from $\langle \delta v_{E \times B,r} \delta J_{\parallel k\text{EM}}^{\text{NA}} \rangle$ and $-\langle \frac{e}{m_e} \delta b_r \delta P_{\parallel k\text{ES}}^{\text{NA}} \rangle$. The last two terms come from kinetic stress $-\langle \frac{e}{m_e} \delta b_r \delta P_{\parallel k\text{EM}}^{\text{NA}} \rangle$ and $-\langle \frac{e}{m_e} \delta b_r \delta P_{\parallel k\text{ES}}^{\text{A}} \rangle$, respectively. The kinetic stress makes dominant contribution to the turbulent flux driven intrinsic current density, but still less than 1% as compared to the BS current density. On the RHS of Eq. (21), the first term represents the non-adiabatic ES contribution $\delta n_{ek\text{ES}}^{\text{NA}}$ to $S_1$. The contribution of adiabatic ES response $\delta n_{ek\text{ES}}^{\text{A}}$ to $S_1$ vanishes naturally as mentioned before. While the second term includes contribution from the non-adiabatic EM density response $\delta n_{ek\text{EM}}^{\text{NA}}$ to $S_1$ and the non-adiabatic ES density response $\delta n_{ek\text{ES}}^{\text{NA}}$ to $S_2$ equally. The third and fourth terms represent the non-adiabatic EM contribution $\delta n_{ek\text{EM}}^{\text{NA}}$ to $S_2$. The last two terms come from the contribution from adiabatic ES density response $\delta n_{ek\text{ES}}^{\text{A}}$ to $S_2$. The intrinsic current density driven by turbulent source is much smaller than that driven by residual turbulent flux. For DIII-D pedestal case, the intrinsic current density can be neglected as compared to the BS current density.

However, examination of Eqs. (18) and (19), the scalings of the ratio between the intrinsic current density driven by turbulence and the BS current density can be written as,

$$\frac{J_{\text{turb}}^{\Gamma}}{J_{\text{BS}}} \propto \frac{T_e^{\frac{3}{4}} T_i}{n_e}, \tag{22}$$

and

$$\frac{J_{\text{turb}}^{\text{S}}}{J_{\text{BS}}} \propto \frac{T_e T_i}{n_e}. \tag{23}$$

This implies that when the electron and/or ion temperature is high enough, the intrinsic current density driven by turbulence may be considerable. Taking the ITER pedestal parameters [36], $q = 3.6$, $R/L_{Te} = 144$, $R/L_n = 64$, $\hat{s} = 1$, $a = 2$m, $\varepsilon = 0.32$, $R = 6.2$m, $B = 4.85$T, $n_e = 4 \times 10^{19}/\text{m}^3$, $T_e = T_i = 3$keV, $\ln\Lambda = 18.7$, $\nu_{ei} = 1.32 \times 10^4$Hz, $\beta_e = 0.21\%$ and keeping the typical DW turbulence parameters, $k_\theta \rho_s \simeq 0.28$, $k_\perp^2 = 2k_\theta^2$, $\gamma_k/\omega_{*e} = 1/10$, $\sum_k I_k = 5 \times 10^{-4}$, $k_\parallel^2 = \frac{\hat{s}^2}{q^2 R^2}$ and $\omega_{kr} = \frac{\omega_{*e}}{2}$, the ratios of the intrinsic current density to the BS current density can be obtained

$$\frac{J_{\text{turb}}^{\Gamma}}{J_{\text{BS}}} = \mp(14.3\% - 24.3\% + 11.5\% - 66.1\%), \tag{24}$$

and

$$\frac{J_{\text{turb}}^{\text{S}}}{J_{\text{BS}}} = -7.8\% + 13.3\% + 0.4\% - 6.3\% + (5.1\% - 3.3\%). \tag{25}$$

For this case, the intrinsic current density driven by turbulent flux can reach about 66% of the BS current density, and thus may be important for modification of the local current density profile. But, the turbulent source driven current density is less than 2% of the local BS current density in pedestal region, and can be neglected. The flux driven intrinsic current is mainly come from the kinetic stress due to adiabatic ES response of parallel electron pressure. The contributions from the non-adiabatic responses are all small. This can be explained from Eq. (12). For ITER pedestal parameters, Eq. (12) becomes to be $\delta\hat{\phi}_k - 2\xi_e \delta\hat{A}_{\parallel k} = (0.148 + i0.266)\delta\hat{\phi}_k$. This indicates that non-adiabatic ES and EM responses of $\delta n_e$, $\delta J_\parallel$ and $\delta P_\parallel$ i.e., Eqs. (3), (5), and (7) become strongly cancelled, and so the non-adiabatic contribution to intrinsic current drive is little. The non-adiabatic contributions to the intrinsic current drive are significantly reduced if the EM effects are important. This is very different from previous works without careful treatment of EM effects. From Eq. (11), $\delta\hat{A}_{\parallel k}$ will flip sign when ITG turbulence is considered. However, the non-adiabatic response is proportional to $\delta\hat{\phi}_k - 2\xi_e \delta\hat{A}_{\|k}$, and $\xi_e = \frac{\omega_{kr}}{k_\| v_{the}}$ will also flip sign for ITG case. Therefore, the non-adiabatic EM response $\sim 2\xi_e \delta\hat{A}_{\|k}$ will not flip sign. This implies that the EM effects still contribute subtraction to the non-adiabatic response, and reduction of the intrinsic current drive due to EM effects could be also expected for ITG case. The results of intrinsic current density based on DIII-D and ITER pedeatal parameters are summarized in Table. 1.

**Table 1.** Results of the estimation for intrinsic current density driven by EM electron DW turbulence for typical pedestal parameters on DIII-D and ITER.

| Ratio of intrinsic current density to BS current density | ES contribution | | EM contribution | |
|---|---|---|---|---|
| | DIII-D | ITER | DIII-D | ITER |
| $\langle \delta v_{E\times B,r} \delta J_\|^{NA} \rangle$ | $\mp 0.58\%$ | $\mp 14.3\%$ | $\pm 0.34\%$ | $\pm 14.3\%$ |
| $\langle -\frac{e}{m_e} \delta b_r \delta P_\|^A \rangle$ | | | $\pm 4.06\%$ | $\pm 66.1\%$ |
| $\langle -\frac{e}{m_e} \delta b_r \delta P_\|^{NA} \rangle$ | | | 0 | $\mp 1.46\%$ |
| Total turbulent flux | $\mp 0.58\%$ | $\mp 14.3\%$ | $\pm 4.40\%$ | $\pm 78.9\%$ |
| $\langle -\frac{e^2}{m_e} \hat{b}\cdot\nabla\delta\phi \delta n_e^{NA} \rangle$ | $-0.52\%$ | $-7.81\%$ | $0.30\%$ | $7.84\%$ |
| $\langle -\frac{e^2}{cm_e} \frac{\partial \delta A_\|}{\partial t} \delta n_e^A \rangle$ | | | $0.38\%$ | $1.81\%$ |
| $\langle -\frac{e^2}{cm_e} \frac{\partial \delta A_\|}{\partial t} \delta n_e^{NA} \rangle$ | | | $4.16 \times 10^{-4}$ | $-0.38\%$ |
| Total turbulent source | $-0.52\%$ | $-7.81\%$ | $0.72\%$ | $9.27\%$ |

## 3. Summary

In this work, it is found that the EM effects on electron DW turbulence driven intrinsic current is significant for ITER pedestal parameters. The non-adiabatic ES response of $\delta n_e$, $\delta J_\|$ and $\delta P_\|$ is strongly cancelled by the non-adiabatic EM response, which results in the contribution from non-adiabatic response to the intrinsic current drive is little. Only the kinetic stress associated with correlation between the radial magnetic fluctuation and the adiabatic ES response of $\delta P_\parallel$ dominates the intrinsic current drive.

Based on DIII-D pedestal parameters, the intrinsic current density driven by the residual turbulent flux and turbulent source is less than 2% of the local BS current density. However, by examination of the scaling of the intrinsic current density driven by turbulence with the electron temperature, the EM electron DW turbulence driven current may become considerable as compared to the BS current density for high electron temperature case. Based on ITER pedestal parameters, the local current density driven by EM electron DW turbulence can reach about 66% of the local BS current density due to the much lower collisionality. Therefore, we conclude that in the high $\beta$ fusion devices like ITER, the EM electron DW turbulence driven intrinsic current density may be important for the modification of the local current density profile in pedestal region and the subsequent ELM control.

Finally, we discuss limitations of the model adopted in this work. The diffusive and convective components of the turbulent current flux are not calculated in the present work, which may be also important for accurate prediction of current density profile and thus the control of ELM in the pedestal region. Therefore, the effects of diffusion and convection of the turbulent current flux induced by EM turbulence on current density profile may be investigated in future. It should be also mentioned that although the symmetry breaking caused by the equilibrium radial electric field $E_\mathrm{r}$ shear has been considered, effects of Doppler shift, $\boldsymbol{E} \times \boldsymbol{B}$ shear suppression of turbulence are absent. Self-consistent including the equilibrium $\boldsymbol{E} \times \boldsymbol{B}$ effects in the study of EM electron DW turbulence driven current is also worth exploring.

## Acknowledgments

We thank P. H. Diamond, J. Q. Li, W. X. Wang, Z. Lin and W. X. Guo for useful discussions. This work was supported by the National Key R&D Program of China under Grant No. 2017YFE0302000, the NSFC under Grant No. 11675059 and the Fundamental Research Funds for the Central Universities, HUST: 2019kfyXMBZ034.

## Appendix. Calculations of intrinsic current drive

The turbulent current flux is consist of two terms. The Reynolds stress-like term can be calculated using Eq. (5),

$$\langle \delta v^*_{E\times B,r} \delta J_\| \rangle = \frac{1}{2}\sum_k \sqrt{\pi} k_\theta \rho_e e n_e v^2_{\text{the}} \xi_e \exp(-\xi_e^2) \left\{ \frac{\omega_{kr}}{|k_\||v_{\text{the}}} \right.$$

$$\left. - \frac{\omega_{*e}}{|k_\||v_{\text{the}}}\left[1 + \eta_e\left(\xi_e^2 - \frac{1}{2}\right)\right]\right\}\left(|\delta\hat{\phi}_k|^2 - 2\xi_e \text{Re}\langle \delta\hat{A}_{\|k}\delta\hat{\phi}_{-k}\rangle\right)$$

$$+ \sum_k \sqrt{\pi} k_\theta \rho_e e n_e v^2_{\text{the}} \xi_e^2 \exp(-\xi_e^2) \frac{\gamma_k}{|k_\||v_{\text{the}}} \text{Im}\langle \delta\hat{A}_{\|k}\delta\hat{\phi}_{-k}\rangle.$$

(A1)

The kinetic stress-like term can be calculated using Eq. (7),

$$\langle -\frac{e}{m_e}\delta P^*_\| \delta b_r \rangle = -\sum_k \sqrt{\pi} k_\theta \rho_e e n_e v^2_{\text{the}} \xi_e^2 \exp(-\xi_e^2) \frac{\gamma_k}{|k_\||v_{\text{the}}} \text{Im}\langle \delta\hat{A}_{\|k}\delta\hat{\phi}_{-k}\rangle$$

$$- \sum_k \sqrt{\pi} k_\theta \rho_e e n_e v^2_{\text{the}} \xi_e^2 \exp(-\xi_e^2) \left\{ \frac{\omega_{kr}}{|k_\||v_{\text{the}}} \right.$$

$$\left. - \frac{\omega_{*e}}{|k_\||v_{\text{the}}}\left[1 + \eta_e\left(\xi_e^2 - \frac{1}{2}\right)\right]\right\}\left[\text{Re}\langle \delta\hat{A}_{\|k}\delta\hat{\phi}_{-k}\rangle - 2\xi_e |\delta\hat{A}_{\|k}|^2\right] -$$

$$+ \sum_k \frac{1}{2} k_\theta \rho_e e n_e v^2_{\text{the}} \text{Im}\langle \delta\hat{A}_{\|k}\delta\hat{\phi}_{-k}\rangle.$$

(A2)

The turbulent source also includes two components. The turbulent source driven by ES field can be written as

$$S_1 = \sum_k -\frac{\sqrt{\pi}}{2} e n_e v_{\text{the}}^2 \exp(-\xi_e^2) \left\{ \frac{\omega_{kr}}{|k_\parallel| v_{\text{the}}} - \frac{\omega_{*e}}{|k_\parallel| v_{\text{the}}} \left[1 + \eta_e \left(\xi_e^2 - \frac{1}{2}\right)\right]\right\} \langle k_\parallel |\delta\hat{\phi}_k|^2 \rangle$$

$$- \sum_k \sqrt{\pi} e n_e v_{\text{the}} \exp(-\xi_e^2) \frac{\omega_{kr} \gamma_k}{|k_\parallel| v_{\text{the}}} \text{Im}\langle \delta\hat{A}_{\parallel k} \delta\hat{\phi}_{-k} \rangle$$

$$+ \sum_k \sqrt{\pi} e n_e v_{\text{the}} \exp(-\xi_e^2) \omega_{rk} \left\{ \frac{\omega_{kr}}{|k_\parallel| v_{\text{the}}} \right.$$

$$\left. - \frac{\omega_{*e}}{|k_\parallel| v_{\text{the}}} \left[1 + \eta_e \left(\xi_e^2 - \frac{1}{2}\right)\right]\right\} \text{Re}\langle \delta\hat{A}_{\parallel k} \delta\hat{\phi}_{-k} \rangle. \tag{A3}$$

The parallel inductive electric field driven source is

$$S_2 = -\sum_k e n_e v_{\text{the}} \left[\omega_{kr} \text{Im}\langle \delta\hat{A}_{\parallel k} \delta\hat{\phi}_{-k} \rangle + \gamma_k \text{Re}\langle \delta\hat{A}_{\parallel k} \delta\hat{\phi}_{-k} \rangle\right]$$

$$+ \sum_k \sqrt{\pi} e n_e v_{\text{the}} \exp(-\xi_e^2) \frac{\gamma_k}{|k_\parallel| v_{\text{the}}} \left[\omega_{kr} \text{Im}\langle \delta\hat{A}_{\parallel k} \delta\hat{\phi}_{-k} \rangle\right.$$

$$\left. + \gamma_k \text{Re}\langle \delta\hat{A}_{\parallel k} \delta\hat{\phi}_{-k} \rangle - 2\xi_e \gamma_k |\delta\hat{A}_{\parallel k}|^2\right]$$

$$+ \sum_k \sqrt{\pi} e n_e v_{\text{the}} \exp(-\xi_e^2) \left\{ \frac{\omega_{kr}}{|k_\parallel| v_{\text{the}}} \right.$$

$$\left. - \frac{\omega_{*e}}{|k_\parallel| v_{\text{the}}} \left[1 + \eta_e \left(\xi_e^2 - \frac{1}{2}\right)\right]\right\} \left[\omega_{kr} \text{Re}\langle \delta\hat{A}_{\parallel k} \delta\hat{\phi}_{-k} \rangle\right.$$

$$\left. - \gamma_k \text{Im}\langle \delta\hat{A}_{\parallel k} \delta\hat{\phi}_{-k} \rangle - 2\xi_e \omega_{kr} |\delta\hat{A}_{\parallel k}|^2\right]. \tag{A4}$$